\documentclass[epjST]{svjour}
\usepackage{graphics}
\begin{document}
\title{Full counting statistics of information content}
\author{Yasuhiro Utsumi\fnmsep\thanks{\email{utsumi@phen.mie-u.ac.jp}}}
\institute{Department of Physics Engineering, Faculty of Engineering, Mie University, Tsu, Mie 514-8507, Japan}
\abstract{
We review connections between the cumulant generating function of full counting statistics of particle number and the R\'enyi entanglement entropy. 
We calculate these quantities based on the fermionic and bosonic path-integral defined on multiple Keldysh contours. 
We relate the R\'enyi entropy with the information generating function, from which the probability distribution function of self-information is obtained in the nonequilibrium steady state. 
By exploiting the distribution, we analyze the information content carried by a single bosonic particle through a narrow-band quantum communication channel. 
The ratio of the self-information content to the number of bosons fluctuates. 
For a small boson occupation number, the average and the fluctuation of the ratio are enhanced. 
} 
\maketitle
\section{Introduction}
\label{intro}

\begin{figure}[ht]
\begin{center}
\resizebox{1.0\columnwidth}{!}{ \includegraphics{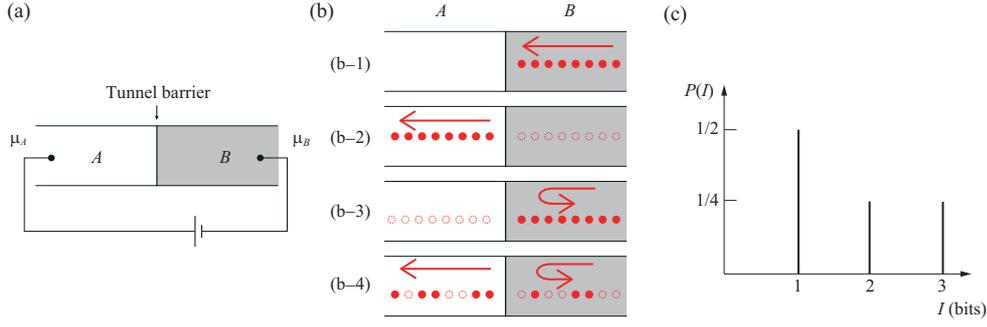} }
\end{center}
\caption{(a) Bipartition of a tunnel junction into subsystems $A$ and $B$. 
(b) Electron scattering processes at the interface between subsystems $A$ and $B$. 
(c) Probability distribution function of self-information. 
}
\label{fig:setups}
\end{figure}

Measurements of the average current and its fluctuation (noise) have been powerful tools to study the quantum transport in mesoscopic systems~\cite{Blanter2000}. 
The probability distribution of current can be treated by the theory of full counting statistics~\cite{Levitov1996,Nazarov2003,BUGS2006}. 
Suppose we partition a mesoscopic conductor, i.e., a tunnel junction, into a subsystem $A$ and a subsystem $B$ [Fig.~\ref{fig:setups} (a)]. 
By applying bias voltage, electrons flow from subsystem $B$ to subsystem $A$. 
The theory of full counting statistics offers a method of calculating the probability distribution function of the number of electrons in subsystem $A$, $P_\tau(N_A)$ at a given measurement time $\tau$. 
It is often convenient to introduce the Fourier transform of the probability distribution function~\cite{Hogg2005}, the characteristic function, 
${\mathcal Z}_\tau(e^{i \chi})=\sum_{N_A} P_\tau(N_A) e^{i N_A \chi}$ 
or its logarithm, the cumulant generating function, 
which yields quantities characterizing the profile of the probability distribution function, $k$th moments 
$\langle N_A^k \rangle = \left. \partial_{i \chi}^k {\mathcal Z}_\tau (e^{i \chi}) \right|_{i \chi=0}$
or cumulant 
$\langle \! \langle N_A^k \rangle \! \rangle = \left. \partial_{i \chi}^k \ln {\mathcal Z}_\tau (e^{i \chi}) \right|_{i \chi=0}$. 

The two subsystems can get entangled after exchanging electrons~\cite{Bennakker2006}. 
The amount of the entanglement between subsystems $A$ and $B$ can be quantified by exploiting the entropy~\cite{Shannon1948,Cover2006,NC2000}. 
It is the von Neumann entropy~\cite{NC2000} $S(\hat{\rho}_A)=-{\rm Tr}_A \left[ \hat{\rho}_A \ln \hat{\rho}_A \right]$ associated to the reduced density matrix obtained after tracing out degrees of freedom of subsystem $B$; 
$\hat{\rho}_A = {\rm Tr}_B \hat{\rho}$. 
In other words, it is the average of the operator of self-information, or the entanglement Hamiltonian, 
$\hat{I}_A = - \ln \hat{\rho}_A$. 
The full counting statistics and the entanglement entropy are related to each other~\cite{Klich2009,FrancisSong2012,Suesstrunk2012,Calabrese2012}. 
In Ref.~\cite{Klich2009}, a nontrivial relation between the current cumulants and the {\it dynamical} entanglement entropy~\cite{Hsu2009} was demonstrated. 
The entropy quantifying the entanglement can be expressed by using the current cumulants of even order as~\cite{Klich2009}, 
\begin{eqnarray}
\langle \! \langle I_A \rangle \! \rangle
=
\sum_{k=1}^\infty 
(-1)^{k+1}(2 \pi)^{2k}
B_{2k}
\langle \! \langle N_A^{2k} \rangle \! \rangle
/(2k)!
\, , 
\label{eqn:KLrelation}
\end{eqnarray}
where $B_k$ are Bernoulli numbers ($B_2=1/6$, $B_4=-1/30$, $B_6=1/42$, $\cdots$). 
The R\'enyi entropy~\cite{Cover2006,Renyi1960} of order $M$, $\ln S_M/(1-M)$, 
where $S_M={\rm Tr}_A \hat{\rho}_A^M$ for quantum cases  (hereafter, we call $S_M$ the R\'enyi entropy~\cite{Nazarov2011})
is another tractable measure of entanglement. 
In Ref.~\cite{FrancisSong2012} the following relation was presented; 
\begin{eqnarray}
\ln S_M = - e^{i \phi} \int_{-\infty}^\infty dz \ln \left[ (1+e^{i \phi} z)^M - e^{i \phi} z^M \right] \mu (z) \, ,  
\label{eqn:fson}
\end{eqnarray}
where the phase is $\phi=0$ for bosons and $\phi=\pi$ for fermions (precisely, the equality for fermions was presented in Ref.~\cite{FrancisSong2012}). 
The spectral density is related to the current cumulant generating function; 
$\mu (z) = e^{-i \phi} \partial_z {\rm Im} \ln {\mathcal Z}_\tau(u=1+e^{-i \phi}/(z+i0))/\pi$. 

In this article, we review these two universal relations based on the multicontour Keldysh technique introduced in Ref.~\cite{Nazarov2011} and developed in Refs.~\cite{Ansari2015,Ansari2017,YU2015,YU2017}. 
In those previous works, the operator representation was adopted. 
Here we introduce path-integral representation. 
We also present another universal relation between the R\'enyi entropy of order $M$, where $M$ is a positive integer, and the current cumulant generating function~\cite{YU2015,YU2017}; 
\begin{eqnarray}
\ln S_M=\sum_{\ell=0}^{M-1} \ln {\mathcal Z}_\tau(e^{i \chi_\ell})
\, , 
\;\;\;\;
\chi_\ell = \frac{2 \pi \ell + \phi}{M}-\phi
\, . 
\label{eqn:renyi_cgf}
\end{eqnarray}
A similar relation connecting the R\'enyi entanglement entropy and the partition function was derived before~\cite{Casini2009}. 
%
In the present article, we focus on the dynamical R\'enyi entanglement entropy in the nonequilibrium steady state realized in the limit of $\tau \to \infty$. 
We also calculate the probability distribution of the self-information and, to illustrate its usage, analyze the information transmission through a narrow-band bosonic quantum communication channel~\cite{Shannon1948,Cover2006,Yamamoto1986}. 

The structure of the paper is as follows. 
In Sec.~\ref{sec:fcs}, we review the full counting statistics for fermions and bosons. 
In Sec.~\ref{sec:fcs_i}, we introduce the self-information of the information theory and explain that it is a random variable. 
Then in Sec.~\ref{sec:path_integ}, we explain the path-integral approach to the full counting statistics and the R\'enyi entanglement entropy. 
In Sec.~\ref{sec:cop}, we apply our technique to an information transmission problem carried by a single bosonic particle. 
In Sec.~\ref{sec:summary}, we summarize this article. 
Step-by-step derivations are given in Appendices.

\section{Full counting statistics}
\label{sec:fcs}

\subsection{Classical picture of transmission of particles}
\label{sec:cp_bf_fcs}

{\it Fermions --}
Lets us consider electron transmission through the tunnel junction~[Fig.~\ref{fig:setups} (a)]. 
An electron is injected from subsystem $B$ to subsystem $A$ and is transmitted (reflected) with the probability $q(1)={\mathcal T}$ ($q(0)={\mathcal T}=1-{\mathcal R}$). 
Suppose during a given measurement time $\tau$, $N$ electrons are injected regularly from subsystem $B$~[Fig.~\ref{fig:setups} (b-1)]. 
If we detect the first electron in subsystem $A$, we set $x_1=1$. 
If not, we set $x_1=0$. 
Suppose we obtain $x_n$ for the $n$th electron ($n=1,\cdots,N$). 
Then we write such a sequence of events $x_1 x_2 \cdots x_{N}$, where $x_n \in {\mathcal X} = \{a_1,a_2 \}= \{0,1 \}$, as ${\bf x}$. 
The probability to find this sequence is $q^{N}({\bf x}) = q(x_1) q(x_2) \cdots q(x_{N})$. 
For example, when all electrons are transmitted [Fig.~\ref{fig:setups} (b-2)], 
the probability is $q^{N}(11 \cdots 1) = q(1) q(1) \cdots q(1) = {\mathcal T}^{N}$. 
On the other hand, when all electrons are reflected [Fig.~\ref{fig:setups} (b-3)], $q^{N}(00 \cdots 0) = q(0) q(0) \cdots q(0) = {\mathcal R}^{N}$. 
The above two events are rare. 
In a given sequence ${\bf x}$, $N p_{ {\bf x}} (a) = \sum_{n=1}^{N} \delta_{x_n,a}$ ($a \in {\mathcal X}$) electrons are transmitted (reflected) for $a=1$ ($a=0$) [Fig.~\ref{fig:setups} (b-4)]. 
The probability to find such a sequence is $q^{N}({\bf x})=\prod_{a \in {\mathcal X} } q(a)^{N p_{ {\bf x}} (a)}={\mathcal T}^{N p_{ {\bf x}} (1)} {\mathcal R}^{N p_{ {\bf x}} (0)}$. 
Therefore, the probability that $N_A$ electrons transmit is, 
\begin{eqnarray}
P_\tau(N_A)=\sum_{ {\bf x} \in {\mathcal X}^N } q^{N}({\bf x})
\, 
\delta_{N_A,N p_{ {\bf x}} (1)}= \left(
\begin{array}{c}
{N} \\ N_A
\end{array} 
\right) {\mathcal T}^{N_A} {\mathcal R}^{{N}-N_A}
\, ,
\; \; 
\left( \begin{array}{c}
{N} \\ n
\end{array} \right) = \frac{N!}{n!(N-n)!}
\, . 
\end{eqnarray}
This is the binomial distribution. 
The characteristic function is~\cite{Levitov1996},  
\begin{eqnarray}
{\mathcal Z}_\tau(e^{i \chi}) = 
\sum_{N_A} P_\tau(N_A) e^{i N_A \chi} =
\left( {\mathcal R} + {\mathcal T} e^{i \chi} \right)^{N} \, . 
\label{eqn:fcs_fermion}
\end{eqnarray}
By taking the derivative of its logarithm, we obtain the first cumulant or moment, the average, 
$\langle \! \langle N_A \rangle \! \rangle = \langle N_A \rangle = \sum_{N_A} N_A P_\tau(N_A) = {N} {\mathcal T}$ 
corresponding to the peak position of the binomial distribution. 
The second cumulant, the noise, is 
$\langle \! \langle N_A^2 \rangle \! \rangle = \langle N_A^2 \rangle -\langle N_A \rangle^2 = \sum_{N_A} N_A^2 P_\tau(N_A) -\left( \sum_{N_A} N_A P_\tau(N_A) \right)^2 = {N} {\mathcal T} {\mathcal R}$, 
which corresponds to the width of the distribution. 

{\it Bosons --}
Let us consider a linear bosonic channel operating at frequency $f$ within a narrow bandwidth ${\mathcal B} \ll f$. 
In the duration of the measurement time $\tau$, $N=\tau {\mathcal B}$ modes are allowed. 
Suppose the $n$th mode contains $x_n \in {\mathcal X}$ bosons (${\mathcal X}= \{ a_1,a_2,a_3,\cdots \}= \{ 0,1,2,\cdots \}$). 
Such probability is given by the geometric distribution 
$q_r( x_n) = ( 1-r ) r^{x_n}$, 
where $r = e^{-\beta_B h f}$ and $\beta_B$ is the inverse temperature of subsystem $B$. 
Therefore the probability to find a configuration $x_1 x_2 \cdots x_{N}$ is 
$q^N({\bf x})=q_r (x_1) \cdots q_r(x_N)$. 
Then the probability to find $N_A$ bosons among $N$ modes is given by the negative binomial distribution as, 
\begin{eqnarray}
P_\tau(N_A)
=
\sum_{{\bf x} \in {\mathcal X}^N} q^N({\bf x}) \, \delta_{N_A,\sum_{j=1}^{\infty} (j-1) N p_{ {\bf x} }(a_j)}
=
(1- r)^N r^{N_A}
\left(
\begin{array}{c}
N-1+N_A \\ N-1
\end{array}
\right)
.
\label{eqn:P_boson}
\end{eqnarray}
The characteristic function is, 
\begin{eqnarray}
{\mathcal Z}_\tau(e^{i \chi}) = \left[ n_B^-(hf) - n_B^+(hf) e^{i \chi} \right]^{-N}
\, , \;\;\;\;
n_B^-(hf) = 1+n_B^+(hf)
\, ,
\label{eqn:fcs_boson}
\end{eqnarray}
where 
$n_B^+(\omega)=1/(e^{\beta_B \omega}-1)$ 
is the Bose distribution function. 
The average number and noise are 
$\langle \! \langle N_A \rangle \! \rangle = n_B^+$ 
and 
$\langle \! \langle N_A^2 \rangle \! \rangle = n_B^+ (1+n_B^+)$, 
respectively.

\subsection{Large deviation}

Once the cumulant generating function is obtained, the probability distribution function can be calculated by performing the inverse Fourier transform. 
In the limit of long measurement time $\tau \to \infty$, 
the number of electrons grows as $N_A \propto \tau$. 
Then, within the saddlepoint approximation, 
\begin{eqnarray}
P_\tau(N_A) = \int^\pi_{-\pi} \frac{d \chi}{2 \pi} e^{-i \chi N_A} {\mathcal Z}_\tau(e^{i \chi}) \approx \exp \left[ \min_{i \chi \in {\mathbf R}} \left[ \ln {\mathcal Z}_\tau(e^{i \chi}) - i \chi N_A \right) \right] \, , 
\label{eqn:ift_saddle}
\end{eqnarray}
which is the Legendre-Fenchel transform~\cite{Touchette2009}. 
The probability distribution function is, 
\begin{eqnarray}
\ln P_\tau(N_A)
\approx
-N
\left \{ 
\begin{array}{cc}
\left( 1- \frac{N_A}{N} \right) \ln \frac{1-\frac{N_A}{N}}{ {\mathcal R} } + \frac{N_A}{N} \ln \frac{\frac{N_A}{N}}{{\mathcal T}}
\,  & ( {\rm Fermions})
\\
\left( 1+ \frac{N_A}{N} \right) \ln \frac{1+n^+}{1+\frac{N_A}{N}} + \frac{N_A}{N} \ln \frac{\frac{N_A}{N}}{n^+}
\,  & ( {\rm Bosons}) 
\end{array}
\right. 
=
-N D(p^* \| q)
\, . 
\label{eqn:pdf_n}
\end{eqnarray}
The result is expressed by the relative entropy $D(p \| q)=\sum_{j=1}^{|{\mathcal X}|} p(a_j) \ln p(a_j)/q(a_j)$ ($|{\mathcal X}|$ is the number of elements in the set ${\mathcal X}$), which measures the difference between the distributions $p$ and $q$.  
The classical picture of the full counting statistics in Sec.~\ref{sec:cp_bf_fcs} is an application of the method of types (Chapter 11 of Ref.~\cite{Cover2006}). 
$p_{\bf x}(a)$ is called a {\it type},  and Eq.~(\ref{eqn:pdf_n}) is a consequence of Sanov's theorem (Appendix \ref{sec:sanov}).
The probability distribution $p^*$ is closest to $q$ in relative entropy, 
i.e., it minimizes $D(p \| q)$ subjected to the constraint $N_A=\sum_{j=1}^{|{\mathcal X}|} (j-1) p(a_j)$. 
For fermions, $p^*$ and $q$ are Bernoulli distributions, 
$(p^*(0),p^*(1))=(1-N_A/N,N_A/N)$ 
and 
$(q(0),q(1))=({\mathcal R},{\mathcal T})$. 
For bosons, they are geometric distributions 
$(p^*(0),p^*(1),\cdots)=(q_{1/(1+N/N_A)}(0),q_{1/(1+N/N_A)}(1),\cdots)$
and
$(q(0),q(1),\cdots)= \left( q_r(0), q_r(1), \cdots \right)$. 
When $p^*=q$, equivalently $N_A= \langle \! \langle N_A \rangle \! \rangle$, the peak of the distribution $\ln P_\tau(N_A) \approx 0$ is realized.

\section{Information as a random variable}
\label{sec:fcs_i}

We recall the basics of information theory~\cite{Cover2006,Fano1961}. 
Suppose ${\mathcal X}=\{ a_1, a_2, a_3, a_4 \}$ is a set of four symbols and the probability of occurrence of each symbol is $q(a_1)=1/2$, $q(a_2)=1/4$, and $q(a_3)=q(a_4)=1/8$. 
If the outcome is $a$, the self-information associated with this outcome is,
\begin{eqnarray}
I_A(a) = - \log_2 q(a) \, ({\rm bit}) \, , 
\;\;\;\;  
I_A(a) = - \ln q(a) \, ({\rm nat}) \, . 
\end{eqnarray}
The self-information of each of the four symbols is $I_A(a_1)=1$, $I_A(a_2)=2$, and $I_A(a_3)=I(a_4)=3$ bits. 
The Shannon entropy is the average of self-information~\cite{Cover2006}; 
$H(q) = \sum_{j=1}^{|{\mathcal X}|} q(a_j) I_A(a_j) = - \sum_{j=1}^{|{\mathcal X}|} q(a_j) \ln q(a_j) = (1/2) \times 1 + (1/4) \times 2 + (1/8) \times 3 + (1/8) \times 3 = 7/4$ (bit). 
The measure of information content introduced here is a random variable, in the sense that it is a value associated to a random event. 
Therefore, we can consider the probability distribution of the self-information; see Chapter 2.7 of Ref.~\cite{Fano1961}. 
The probability distribution of self-information (in bits) is, 
$
P(I_A) = \sum_{j=1}^{|{\mathcal X}|} q(a_j) \delta(I_A-I_A(a_j)) = (1/2) \delta(I_A-1) + (1/4) \delta(I_A-2) + 2 \times (1/8) \delta(I_A-3) 
$, 
which is visualized in Fig.~\ref{fig:setups} (c). 
It is clear that the Shannon entropy is the average of this probability distribution function $H=\langle I_A \rangle = \int dI_A P(I_A) I_A$.

Let us consider the transmission of particles in Sec.~\ref{sec:cp_bf_fcs}. 
In the following, we measure the information content in nats . 
A sequence ${\bf x}$ carries the self-information amount to $I_A({\bf x})=-\ln q^N({\bf x})$. 
The probability distribution function of self-information and its Fourier transform, {\it information generating function}~\cite{Golomb1966,Guiasu1985}, are, 
\begin{eqnarray}
P_\tau(I_A)=\sum_{{\bf x}} q^N({\bf x}) \delta(I_A - I_A({\bf x}))
\, , \;
S_{1-i \xi} = \int d I_A e^{i \xi I_A} P_\tau(I_A) 
=\left( \sum_{j=1}^{|{\mathcal X}|} q(a_j)^{1-i \xi} \right)^{N} .
\nonumber \\
\label{eqn:pdfi}
\end{eqnarray}
Since there is an apparent formal similarity, in the following we use the terms ``information generating function" and ``R\'enyi entropy" interchangeably. 
The information generating functions for fermions and bosons are, 
\begin{eqnarray}
S_{M} = \left( {\mathcal T}^{M} + {\mathcal R}^{M} \right)^{N}
\, , \;\;
S_{M} = \left[ {n_B^-}(hf)^M - {n_B^+}(hf)^M \right]^{-N}
\, .
\label{eqn:igf_f_b}
\end{eqnarray}
The average is the Shannon entropy, 
$\langle \! \langle I_A \rangle \! \rangle = \left. \partial_{i \xi} \ln S_{1-i \xi} \right|_{i \xi=0}=N H(q)$
where 
$H(q)=H_2( {\mathcal T} )=-{\mathcal T} \ln {\mathcal T}-(1-{\mathcal T}) \ln (1-{\mathcal T})$ 
is the binary entropy for fermions and 
$H(q)=H_g(n_B^+)= (1+n_B^+) \ln (1+n_B^+) - n_B^+ \ln n_B^+$
is the entropy for bosons. 
Similar to Eq.~(\ref{eqn:pdf_n}), the probability distribution is calculated as, 
\begin{eqnarray}
\ln P_\tau(I_A) + I_A \approx 
N H_2 \left( \frac{I_A/N+\ln {\mathcal T} }{ \ln ({\mathcal T}/{\mathcal R}) } \right)
\, ,
\;\;
N H_g \left( \frac{ I_A/N +\ln (1-r)}{-\ln r} \right)
\, .
\label{eqn:pdf_I}
\end{eqnarray}

Let us rewrite Eq.~(\ref{eqn:pdfi}) and check its meaning. 
For $N \gg 1$, typical sequences are around the peak position $\langle \! \langle I_A \rangle \! \rangle =N H(q)$ of the distribution function $P_\tau(I_A)$. 
For $\epsilon>0$, the intensity around the peak 
$ | I_A/N - H(q)| \leq \epsilon$ is the probability to find typical sequences; 
$\int^{N (H(q)+\epsilon)}_{N (H(q)-\epsilon)} dI_A P_\tau(I_A) = \sum_{{\bf x} \in T(N,\epsilon)} q^N({\bf x})$
where $T(N,\epsilon)$ is a set of all $\epsilon$-typical sequences of length $N$~\cite{NC2000} satisfying 
$e^{-N(H(q)+\epsilon)} \leq  q^N({\bf x}) \leq e^{-N(H(q)-\epsilon)}$.

\section{Path-integral approach} 
\label{sec:path_integ}

{\it Full counting statistics --}
The Hamiltonian of the tunnel junction is given by, 
$\hat{H} = \hat{H}_A + \hat{H}_B + \hat{V}$. 
The Hamiltonians for the leads $r=A,B$ are, 
$\hat{H}_r = \sum_k \epsilon_{r \, k} \hat{c}_{rk}^\dagger \hat{c}_{rk}$, 
where $\hat{c}_{rk}$ annihilates a particle in quantum state $k$ in the lead $r$. 
The tunnel Hamiltonian is 
$\hat{V} = J \sum_{k,k'}( \hat{c}_{Ak}^\dagger \hat{c}_{Bk'} + \hat{c}_{Bk'}^\dagger \hat{c}_{Ak})$. 
Then the probability distribution function of a particle number in the subsystem $A$ and its characteristic function are 
\begin{eqnarray}
P_\tau(N_A)={\rm Tr}_A \left[ \hat{\Pi}_{N_A}
\hat{\rho}_A (\tau) \right] \, ,
\;\;\;
{\mathcal Z}_\tau(e^{i \chi}) = {\rm Tr}_A \left[e^{i \chi \hat{N}_A} \hat{\rho}_A (\tau) \right]
\, ,
\label{eqn:pdfnq}
\end{eqnarray}
where $\hat{N}_A=\sum_k \hat{c}_{Ak}^\dagger \hat{c}_{Ak}$ is the operator of the number of particles in subsystem $A$ and $\hat{\Pi}_{N_A}=\int_{-\pi}^\pi d \chi e^{i (\hat{N}_A-N_A) \chi}/(2 \pi)$ is the projection operator. 
The phase $\chi$ is called the {\it counting field}. 
The reduced density matrix of subsystem $A$ at time $\tau$ is prepared by the following protocol. 
Initially at time $t=0$, the subsystems are decoupled and each subsystem is equilibrated with the inverse temperature $\beta_{A(B)}$ and the chemical potential $\mu_{A(B)}$. 
The equilibrium density matrix of the subsystem $r$ is, 
$\hat{\rho}_{{\rm eq} \, r} = e^{-\beta_{r} (\hat{H}_{r} - \mu_{r} \hat{N}_{r})} /Z_r$, 
where the equilibrium partition function is 
$Z_r=Z(\beta_r,\mu_r)={\rm Tr}_r \left[ e^{-\beta_{r} (\hat{H}_{r} - \mu_{r} \hat{N}_{r})} \right]$. 
The explicit form is, 
$\ln Z_r= -e^{i \phi} \int d \omega \rho_r(\omega) \ln \left( 1-e^{i \phi -\beta_r (\omega - \mu_r)} \right)$, 
where
$\rho_r(\omega) = \sum_k \delta (\omega - \epsilon_{rk})$
is the DOS of particles in the lead $r$. 
At $t=0$, we switch on the coupling $\hat{V}$ and let the total system evolve till $t=\tau$. 
Then the reduced density matrix of the subsystem $A$ at $t=\tau$ is, 
\begin{eqnarray}
\hat{\rho}_A (\tau) = {\rm Tr}_B \left[ e^{-i \hat{H} \tau} \hat{\rho}_{{\rm eq} \, A} \hat{\rho}_{{\rm eq} \, B} e^{i \hat{H} \tau} \right]
\, . 
\label{eqn:rdm}
\end{eqnarray}

\begin{figure}
\begin{center}
\resizebox{1.0\columnwidth}{!}{ \includegraphics{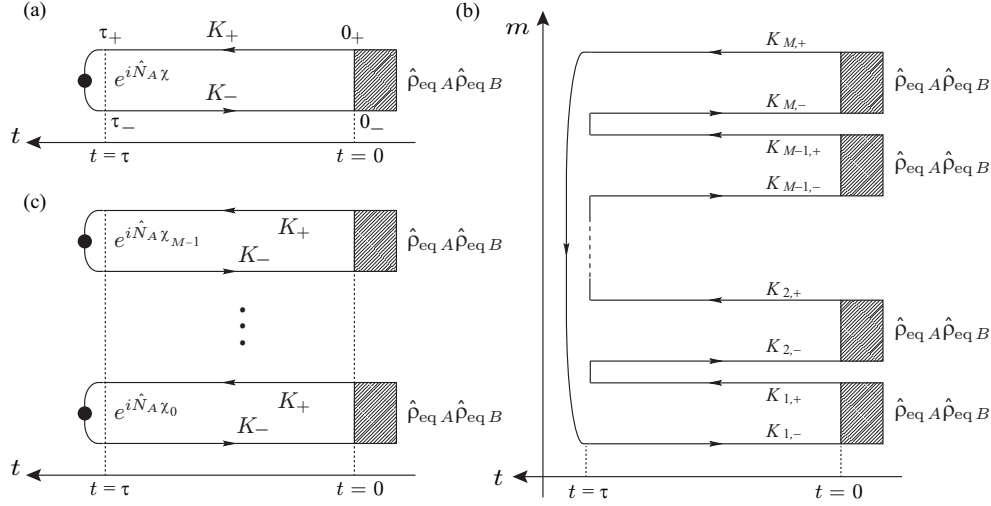} }
\end{center}
\caption{(a) Keldysh contour $K$ consisting of the forward (upper) branch $K_+$ and the backward (lower) branch $K_-$. 
The shaded box is the initial density matrix. 
The solid dot at $t=\tau$ indicates the operator $e^{i \hat{N}_A \chi}$. 
(b) The sequence of $M$ Keldysh contours. 
Time $\tau$ on the upper branch of $m$th Keldysh contour $K_{m,+}$ is connected to time $\tau$ on the lower branch of $m+1$th Keldysh contour $K_{m+1,-}$ ($m=1,\cdots M$ and $K_{M+1,-}=K_{1,-}$). 
(c) $M$ disconnected Keldysh contours obtained after the discrete Fourier transform. 
Solid dots at $t=\tau$ indicate operators $e^{i \hat{N}_A \chi_\ell}$ ($\ell=0, \cdots , M-1$). 
}
\label{fig:KeldyshCs}
\end{figure}

We introduce the Keldysh path-integral~\cite{Schoen1990,Weiss2012,Kamenev2011} representation of the characteristic function 
${\mathcal Z}_\tau(e^{i \chi}) = \int {\mathcal D}[c_{rk}(t)^* c_{rk}(t)] e^{i {\mathcal S}(\chi)}/(Z_A Z_B)$ (see Appendix~\ref{sec:tst} for detailed derivations), 
where the action is, 
\begin{eqnarray}
{\mathcal S}(\chi) = {\mathcal S}_K( \{ c_{rk}(t)^*, c_{rk}(t) \} ) +  \sum_{r=A,B}{\mathcal S}_{\rm b.c.} (c_{rk}(\tau_\pm)^*, c_{rk}(\tau_\pm);{\lambda_r})
\, . 
\label{eqn:cgf_action}
\end{eqnarray}
Here $c_{rk}$ is a complex number (Grassmann number) and $c_{rk}^*$ is its complex conjugation (conjugation) for a bosonic (fermionic) field~\cite{Negele1988}. 
The action is defined on the Keldysh contour $K$ [Fig.~\ref{fig:KeldyshCs} (a)], 
\begin{eqnarray}
i {\mathcal S}_K &=& \sum_{\sigma = \pm} i \sigma \int_0^\tau dt_\sigma \biggl[ \sum_{rk} c_{rk}(t_\sigma)^* (i \partial_{t_\sigma} -\epsilon_{rk}) c_{rk}(t_\sigma) - J \sum_{k,k'} (c_{Ak} (t_\sigma)^* c_{Bk'}(t_\sigma) \nonumber \\ && 
+ c_{Bk'} (t_\sigma)^* c_{Ak}(t_\sigma)) \biggl] + \sum_{r k} c_{rk}(0_+)^* (c_{rk}(0_-) e^{-\beta_r (\epsilon_r - \mu_r)}-c_{rk}(0_+))
\, , 
\label{eqn:action_bulk}
\end{eqnarray}
where time $t_\pm$ is defined on $K_\pm$. 
The first term is defined on the forward (upper) and backward (lower) branches $K_+$ and $K_-$. 
The second term imposes the boundary condition at $t=0_+$ and $t=0_-$. 
The action determining the boundary condition at $t=\tau_+$ and $t=\tau_-$ is, 
\begin{eqnarray}
i {\mathcal S}_{\rm b.c.} (c_{rk}(\tau_\pm)^*, c_{rk}(\tau_\pm);{\lambda_r})
=
\sum_{k} c_{rk}(\tau_-)^* (c_{rk}(\tau_+) e^{i \lambda_r}-c_{rk}(\tau_-))
\, , 
\label{eqn:action_boundary_condition}
\end{eqnarray}
where $\lambda_A=\chi+\phi$ and $\lambda_B=\phi$. 
Then, after the functional integral, and in the limit of long measurement time, we obtain
\begin{eqnarray}
\ln \frac{ {\mathcal Z}_\tau(e^{i \chi}) }{ {\mathcal Z}_0(e^{i \chi}) } & \approx & -e^{i \phi} \frac{\tau}{2 \pi} \int d \omega \ln \det \left[ {\bf \tau}_0 - J^2 {\bf \tau}_3 {\bf g}_{A,\chi+\phi}(\omega) {\bf \tau}_3 {\bf g}_{B,\phi}(\omega) \right] \, ,
\label{eqn:cgf_omega}
\end{eqnarray}
where ${\bf \tau}_0={\rm diag}(1,1)$ is the identity matrix and ${\bf \tau}_3={\rm diag}(1,-1)$ is the Pauli matrix in the Keldysh space. 
The $2 \times 2$ modified Keldysh Green-function matrix (see, e.g.,~\cite{BUGS2006,Esposito2009,UGS2006,SU2008,US2009,Urban2010,Novotny2011,UEUA2013}) is, 
\begin{eqnarray}
{\bf g}_{r, \lambda}(\omega)=-2 \pi i \, \rho_r (\omega)
\left [
\begin{array}{cc}
1/2+n_{r,\lambda}^+(\omega) & n_{r,\lambda}^+(\omega) e^{-i \lambda} \\
n_{r,\lambda}^-(\omega) e^{ i \lambda} & 1/2+n_{r,\lambda}^+(\omega)
\end{array}
\right]
,
\label{eqn:kgm}
\end{eqnarray}
where we neglected the real part. 
We introduced, 
\begin{eqnarray}
{n}_{r,\lambda}^-(\omega)=\frac{1}{1-e^{-\beta_r (\omega - \mu_r)+i \lambda}}
\, , 
\;\;\;\;
n_{r,\lambda}^+(\omega)=e^{-\beta_r (\omega - \mu_r)+i \lambda}
\, 
n_{r,\lambda}^-(\omega)
\, , 
\end{eqnarray}
which is the particle distribution function when $\lambda=\phi$. 
For $\phi=0$, the Bose distribution function 
$n_{r,0}^+=n_r^+=1/(e^{\beta_r (\omega - \mu_r)}-1)$ 
and 
$n_{r,0}^-=n_r^-=1+n_r^+$
are obtained. 
For $\phi=\pi$, the Fermi distribution function 
$n_{r,\pi}^+=-f_r^+=-1/(e^{\beta_r (\omega - \mu_r)}+1)$
and 
$n_{r,\pi}^-=f_r^-=1-f_r^+$
are obtained. 
At $\tau=0$, the two subsystems are decoupled and the characteristic function is, 
$\ln {\mathcal Z}_0(e^{i \chi}) 
= \ln Z(\beta_A,\mu_A+i \chi/\beta_A)/Z(\beta_A,\mu_A)
= -e^{i \phi} \int d \omega \rho_A(\omega) \ln ( n_{A,\phi}^-(\omega) - n_{A,\phi}^+(\omega) e^{i \chi} )$. 

Further calculations lead to: 
\begin{eqnarray}
\ln \frac{ {\mathcal Z}_\tau (e^{i \chi}) }{ {\mathcal Z}_0(e^{i \chi}) } = -e^{i \phi} \frac{\tau}{2 \pi} \int d \omega \ln \frac {\tilde{n}_{A,\phi}^+(\omega) e^{i \chi} - \tilde{n}_{A,\phi}^-(\omega)} {n_{A,\phi}^+(\omega) e^{i \chi} - n_{A,\phi}^-(\omega)} , 
\label{eqn:cgf_long_tau}
\end{eqnarray}
where we omitted a constant to satisfy the normalization condition 
${\mathcal Z}_\tau(1)=1$. 
We introduced the effective particle distribution function, 
$\tilde{n}_{A,\phi}^\pm(\omega) = {\mathcal R}(\omega) {n}_{A,\phi}^\pm(\omega) + {\mathcal T}(\omega) {n}_{B,\phi}^\pm(\omega)$, 
where the transmission and reflection probabilities are 
${\mathcal T}(\omega) = 1-{\mathcal R}(\omega) = 4 \pi^2 J^2 \rho_A(\omega) \rho_B(\omega)/[1+\pi^2 J^2 \rho_A(\omega) \rho_B(\omega)]^2$. 
For fermions, $\phi=\pi$, Eq~(\ref{eqn:fcs_fermion}) is obtained from Eq.~(\ref{eqn:cgf_long_tau}) by taking the zero temperature limit $\beta_A, \beta_B \to \infty$ for the energy-independent transmission probability 
${\mathcal T}(\omega)={\mathcal T}$. 
The number of injected fermions is $N=\tau (\mu_B - \mu_A)/(2 \pi)$ for $\mu_B > \mu_A$. %
For bosons $\phi=0$, Eq.~(\ref{eqn:fcs_boson}) is derived for the narrow-band channel, i.e., the energy filter ${\mathcal T}(\omega)=h {\mathcal B} \, \delta( \omega - h f)$, where the bandwidth ${\mathcal B}$ is much smaller than the signal frequency $f$, when the subsystem $A$ is empty, $n_{A,0}^+(h f)=0$.

{\it R\'enyi entanglement entropy --}
The probability distribution function of self-information and the information generating function are 
\begin{eqnarray}
P_\tau(I_A) = {\rm Tr}_A \left[ \hat{\rho}_A(\tau) \delta (I_A + \ln \hat{\rho}_A(\tau) ) \right]
\, ,
\;\;\;\; 
S_{1-i \xi} = {\rm Tr}_A \left[ \hat{\rho}_A(\tau)^{1-i \xi} \right]
\, . 
\label{eqn:pdfiq}
\end{eqnarray}
The spectrum of the entanglement Hamiltonian, the entanglement spectrum~\cite{Petrescu2014}, is closely related to this distribution;
${\rm Tr}_A \left[ \delta(I_A-\hat{I}_A) \right] = e^{I_A} P_\tau(I_A)$. 
This relation implies $ \langle e^{I_A} \rangle = {\rm rank} \hat{\rho}_A$, 
which is reminiscent of the Jarzynski equality~\cite{Jarzynski1997,Campisi2011}. 
As an example, let us consider the density matrix 
$\hat{\rho}_A= \left ( \sum_{j=1}^{|{\mathcal X}|} q(a_j) |j \rangle \langle j| \right)^{\otimes N}$, where $|j \rangle$ is an orthonormal set. 
Then Eq.~(\ref{eqn:pdfiq}) reduces to the classical case, Eq.~(\ref{eqn:pdfi}). 
By applying Jensen's inequality to the Jarzynski equality $ \langle e^{I_A} \rangle = |{\mathcal X}|^N$, we obtain $ \langle I_A \rangle \leq N \ln |{\mathcal X}|$, 
i.e., $N \ln |{\mathcal X}|$ is the maximum entropy. 

At $\tau =0$, when the subsystems are decoupled, the information generating function is, 
$s_{M} = {\rm Tr}_A \left[ \hat{\rho}_{{\rm eq} A}^{M} \right]=Z(M \beta_A,\mu_A)/Z_A^M$. 
At finite $\tau$, when the coupling induced the correlations between the two subsystems, the information generating function is calculated by the replica trick: 
We first calculate it for a positive integer $M$ and then perform the analytic continuation $M \to 1 - i \xi$. 
The path-integral representation of the R\'enyi entropy of a positive integer $M$ order, 
$S_M={\rm Tr}_A \left[\hat{\rho}_A(\tau) \cdots \hat{\rho}_A(\tau) \right]$, 
is evaluated by extending the contour $K$ to a sequence of $M$ Keldysh contours~\cite{Nazarov2011} [Fig.~\ref{fig:KeldyshCs} (b)]; 
$S_M = (Z_A Z_B)^{-M} \int {\mathcal D} [c_{mrk}(t_m)^* c_{mrk}(t_m)] e^{i {\tilde {\mathcal S} }}$,
where the action is, 
\begin{eqnarray}
{\tilde {\mathcal S}} = \sum_{m=1}^M {\mathcal S}_{K_m} ( \{ c_{mrk}(t_m)^*,c_{mrk}(t_m) \}) + {\mathcal S}_{ {\rm b.c.} }(c_{mBk}(\tau_{m,\pm})^*,c_{mBk}(\tau_{m,\pm});{\phi})+{\mathcal S}_{ {\rm b.c.} A}^m
. 
\nonumber \\
\end{eqnarray}
Here, time $t_{m,\pm}$ is defined on the upper (lower) branch of $m$th Keldysh contour $K_{m,\pm}$. 
The fields $c_{mrk}$ and thus the action ${\mathcal S}_{K_m}$ are defined on the $m$th Keldysh contour $K_{m}$. 
The definition of the action ${\mathcal S}_{K_m}$ is the same as Eq.~(\ref{eqn:action_bulk}). 
The action imposing the boundary condition at $t=\tau_{m,+}$ and $t=\tau_{m,-}$ for the fields of the subsystem $B$, ${\mathcal S}_{ {\rm b.c.} }$, is the same as Eq.~(\ref{eqn:action_boundary_condition}). 
For the fields of the subsystem $A$, the action ${\mathcal S}_{{\rm b.c.} A}^m$ imposes the boundary condition connecting $t=\tau_{m,+}$ on the upper branch of $m$th Keldysh contour and $t=\tau_{m+1,-}$ on the lower branch of $m+1$th Keldysh contour; 
\begin{eqnarray}
i {\mathcal S}_{{\rm b.c.} A}^m = \sum_{k} c_{m+1Ak} (\tau_-)^* (c_{mAk} (\tau_+) - c_{m+1 Ak} (\tau_-)) \, ,
\label{eqn:bcA}
\end{eqnarray}
where $c_{M+1Ak}(\tau_-)=c_{1Ak}(\tau_-)e^{-i \phi}$. 
This action can be diagonalized by the discrete Fourier transform; 
\begin{eqnarray}
c_{mrk}(t) = \frac{1}{\sqrt{M}} \sum_{\ell =0}^{M-1} c_{\ell rk}(t) e^{-i (2 \pi \ell + \phi) m /M} \, ,
\label{eqn:ft}
\end{eqnarray}
which fulfills the periodic or anti-periodic boundary condition 
$c_{m+M rk}(t) = c_{m rk}(t) e^{-i \phi}$. 
The resulting action is, 
\begin{eqnarray}
\sum_{\ell=1}^{M} {\mathcal S}_{{\rm b.c.} A}^m = \sum_{\ell=0}^{M-1} {\mathcal S}_{{\rm b.c.}} (c_{\ell Ak}(\tau_\pm)^*, c_{\ell Ak}(\tau_\pm); \chi_\ell + \phi)
\, , 
\;\;
\chi_\ell = \frac{2 \pi \ell + \phi}{M}-\phi
\, . 
\end{eqnarray}
The discrete Fourier transform introduces a discrete counting field $\chi_\ell$ at $t=\tau$. 
Since the action defined on the Keldysh contour Eq.~(\ref{eqn:action_bulk}) is quadratic, it is diagonal after the Fourier transform; 
$ \sum_{m=1}^{M} {\mathcal S}_{K_m} ( \{ c_{m rk}^*,c_{m rk} \} ) =\sum_{\ell=0}^{M-1} {\mathcal S}_K ( \{ c_{\ell rk}^*,c_{\ell rk} \} ) $. 
The action imposing the boundary condition for subsystem $B$ is also diagonal in $\ell$; 
$\sum_{m=1}^{M} {\mathcal S}_{\rm b.c.} ( c_{m B k}(\tau_{m,\pm})^*, c_{m B k}(\tau_{m,\pm});\phi) = \sum_{\ell=0}^{M-1} {\mathcal S}_{\rm b.c.} ( c_{\ell B k}(\tau_\pm)^*, c_{\ell B k}(\tau_\pm);\phi)$. 
In this way, we separate connected $M$ Keldysh contours [Fig.~\ref{fig:KeldyshCs} (b)] into disconnected $M$ Keldysh contours [Fig.~\ref{fig:KeldyshCs} (c)]. 
The action is expressed by using the action of the cumulant generating function (\ref{eqn:cgf_action}), 
\begin{eqnarray}
{\tilde {\mathcal S}} = \sum_{\ell=0}^{M-1} {\mathcal S}(\chi_\ell) . 
\label{eqn:diag_action}
\end{eqnarray}
By noticing that the Jacobian of the Fourier transform is 1, we obtain the relation~(\ref{eqn:renyi_cgf}). 

\begin{figure}
\begin{center}
\resizebox{0.9\columnwidth}{!}{ \includegraphics{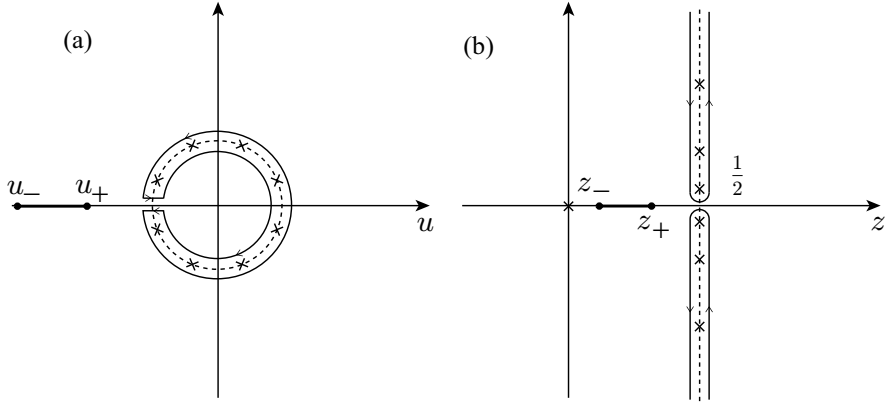} }
\end{center}
\caption{(a) Contour of integral $C$ encircling singularities at $u=e^{i ((2 \pi \ell+\phi)/M-\phi)}$ (in this panel, $\phi=\pi$ and $M=8$) denoted by crosses. 
The dashed line indicates the unit circle. 
There is a branch cut connecting two branch points 
$u_+=\tilde{n}_{A,\phi}^-(\omega)/\tilde{n}_{A,\phi}^+(\omega)$ 
and 
$u_-={n}_{A,\phi}^-(\omega)/{n}_{A,\phi}^+(\omega)$ on the real axis 
(in this panel, we assumed $u_+ > u_-$). 
(b) The contour $C$ and singularities after the variable transform $z=-e^{-i \phi}/(1-u)$. 
The dashed line indicates ${\rm Re} z=-e^{-i \phi}/2$. 
}
\label{cont_integ_f}
\end{figure}

For further calculations, we define $u=e^{i \chi}$ and rewrite the summation over $\ell$ as the contour integral~\cite{Casini2009}; 
\begin{eqnarray}
\ln S_M = \int_C \frac{d u}{2 \pi i} \sum_{\ell=0}^{M-1} \frac{\ln {\mathcal Z}_\tau(u)}{u-e^{i \chi_\ell}}
= 
\int_C \frac{d u}{2 \pi i} \partial_u \ln \left( u^M-e^{i \phi (1-M)} \right) \ln {\mathcal Z}_\tau(u)
\, , \label{eqn:casini}
\end{eqnarray}
where the contour $C$ is taken so that it encircles poles at $u=e^{i \chi_\ell}$ ($\ell=0,\cdots ,M-1$) [Fig.~\ref{cont_integ_f} (a)]. 
Then we substitute the expression for $\tau \to \infty$ Eq.~(\ref{eqn:cgf_long_tau}). 
The integrand has a branch cut on the real axis connecting two branch points 
$u_+=\tilde{n}_{A,\phi}^-(\omega)/\tilde{n}_{A,\phi}^+(\omega)$ 
and 
$u_-={n}_{A,\phi}^-(\omega)/{n}_{A,\phi}^+(\omega)$. 
By a variable transform $z=-e^{-i \phi}/(1-u)$, which transforms a unit circle to a line ${\rm Re} z= -e^{i \phi}/2$, and by noticing that the branch cut stays on the real axis after this transform [Fig.~\ref{cont_integ_f} (b)], Eq.~(\ref{eqn:casini}) can be transformed into Eq.~(\ref{eqn:fson}). 
The explicit form of the spectral density associated to Eq.~(\ref{eqn:cgf_long_tau}) is, 
\begin{eqnarray}
\mu (z) = \int d \omega \left[ \frac{\tau}{2 \pi} \delta(z-e^{-i \phi} \tilde{n}_{A,\phi}^+(\omega)) + \left(\rho_A(\omega) - \frac{\tau}{2 \pi} \right) \delta(z-e^{-i \phi} {n}_{A,\phi}^+(\omega)) \right]
.
\label{eqn:mu_long_tau}
\end{eqnarray}
The first term in the square brackets contains the effective distribution function and thus is related to particle transmission and reflection. 
The second term is the bulk thermodynamic entropy of subsystem $A$ and the overcounting term. 
In the limit of zero temperature, Eq.~(\ref{eqn:mu_long_tau}) is the effective-transparency density~\cite{Abanov2009} and the above discussions are applicable to a finite $\tau$ case since, for noninteracting fermions, singularities are always on the negative real axis of complex $u$-plane~\cite{Abanov2008}. 

The information generating functions in Eq~(\ref{eqn:igf_f_b}) are obtained by substituting Eq.~(\ref{eqn:mu_long_tau}) into Eq.~(\ref{eqn:fson}). 
For fermions $\phi=\pi$, again we take the zero temperature limit $\beta_A, \beta_B \to \infty$ and consider the energy-independent transmission probability. 
For bosons $\phi=0$, we set ${\mathcal T}(\omega)=h {\mathcal B} \, \delta( \omega - h f)$ and $n_{A,0}^+(h f)=0$. 

The relation (\ref{eqn:KLrelation}) is obtained by expanding the RHS of Eq.~(\ref{eqn:renyi_cgf}) in powers of $\chi_\ell$ and then performing the summation over $\ell$; 
\begin{eqnarray}
\ln S_M = \sum_{k=1}^\infty \frac{ \langle \! \langle N_A^{k} \rangle \! \rangle }{k!} \left( \frac{2 \pi i}{M} \right)^k \times 
\left \{
\begin{array}{cc}
\zeta(-k,1) - \zeta(-k,M) & (\phi=0) \\
 \zeta(-k,\frac{1-M}{2}) - \zeta(-k,\frac{1+M}{2}) & (\phi=\pi) 
\end{array}
\right. 
.
\end{eqnarray}
Here 
$\zeta(s,a)=\sum_{\ell=0}^\infty(a+\ell)^{-s}$ is the Hurwitz zeta function. 
By the analytic continuation $M \to 1-i \xi$ and by differentiating in $\xi$, we obtain Eq.~(\ref{eqn:KLrelation}) except for the imaginary number $-i \pi \langle \! \langle N_A \rangle \! \rangle$ for bosons, which may be an artifact and is neglected.

\section{Information transmission through a bosonic quantum channel}
\label{sec:cop}

It is known that noninteracting bosons cannot be entangled if the initial decoupled systems are in equilibrium~\cite{Bennakker2006,Xiang-bin2002}. 
Therefore, our average self-information does not measure the amount of entanglement for bosons. 
Here we present that the probability distribution function of self-information can be used to analyze the performance of a quantum communication channel~\cite{Cover2006} by regarding subsystem $A$ as a receiver side and subsystem $B$ as a transmitter side. 
Let us consider a single narrow-band bosonic channel. 
The transmitter side generates signals, the thermal noise of temperature $\beta_B^{-1}$, with average power ${\mathcal P} _A=hf {\mathcal B} n_B^+$. 
We set $\beta_A \to \infty$ to suppress the detector noise. 
Then we ask a question addressed in Ref.~~\cite{Yamamoto1986}: 
{\it How much information can be transmitted by a single boson?}
The quantity we consider is the ratio of the self-information content to the number of bosonic particles, $\eta =I_A/N_A$. 
It is a random variable, since both the numerator and the denominator are random variables. 
It is also an analog of the efficiency, the ratio of output to input, 
whose fluctuations have been addressed recently~\cite{Verley2014,Polettini2015,Proesmans2016,Okada2017}. 
In the limit of long measurement time $\tau \to \infty$, 
the average approaches $ \langle \eta \rangle \approx \langle I_A \rangle/\langle N_A \rangle = H_g(n_B^+)/n_B^+$ as predicted in Ref.~\cite{Yamamoto1986}. 
	
In order to analyze the probability distribution of $\eta$, we introduce the self-information associated with a state after the projective measurement of the particle number $N_A$ in the subsystem $A$; 
$\hat{I}_A^\prime = - \ln \hat{\rho}_{A}^\prime$, 
where 
$\hat{\rho}_{A}^\prime = \sum_{N_A} \hat{\Pi}_{N_A} \hat{\rho}_A \hat{\Pi}_{N_A}$.
The joint probability distribution of $I_A^\prime$ and $N_A$ is
$P_\tau(I_A^\prime, N_A) = {\rm Tr}_A \left[ \hat{\Pi}_{N_A} \hat{\rho}_A \hat{\Pi}_{N_A} \delta \left( I_A^\prime + \ln \hat{\rho}_{A}^\prime \right) \right]$. 
The information generating function is, 
\begin{eqnarray}
S_{1-i \xi}(\chi) = \int d I_A^\prime \sum_{N_A} e^{i N_A \chi + i I_A^\prime \xi} P_\tau(I_A^\prime, N_A) = {\rm Tr}_A \left[ \left( e^{i \hat{N}_A \chi/(1-i \xi)} \hat{\rho}_A \right)^{1-i \xi} \right],
\label{eqn:mrenyi}
\end{eqnarray}
where we used the {\it local particle number super-selection}~\cite{Bennakker2006,YU2017,Wiseman2003,Klich2009_2}, which ensures $[\hat{\rho}_A,\hat{N}_A]=0$~\cite{YU2017}. 
Then for the narrow-band channel when the detector noise is absent $\beta_A \to \infty$, the characteristic function is 
$\ln S_M(\chi)=-N \ln [n_B^-(h f)^M - e^{i \chi} n_B^+(hf)^M]$. 
The joint probability distribution is then, 
$P(I_A^\prime,N_A) = P_\tau(N_A) \delta \left( I_A^\prime + \ln (1-r)^N r^{N_A} \right)$, 
where $P_\tau(N_A)$ is given by Eq.~(\ref{eqn:P_boson}). 
The probability distribution is, 
\begin{eqnarray}
P_\tau(\eta) = \sum_{N_A=1}^\infty \int d I_A^\prime P_\tau(I_A^\prime,N_A) \delta(\eta - I_A^\prime/N_A) \approx |N_A^*|^2 \, P(N_A^*)/(\eta + \ln r)
\, ,
\end{eqnarray}
where 
$N_A^*/N = -\ln (1-r)/(\eta + \ln r)$. 
From the condition $N_A^* \geq 0$, we observe that the fluctuation of $\eta$ is bounded below 
$\eta > - \ln r = \beta_B h f$. 
In the limit of $\tau \to \infty$, we can adopt Eq. (\ref{eqn:pdf_n}) and see that the peak position is at $1/(1+N/N_A^*)=r$ equivalently $\eta = \langle \eta \rangle$.

\begin{figure}
\begin{center}
\resizebox{1.0\columnwidth}{!}{ \includegraphics{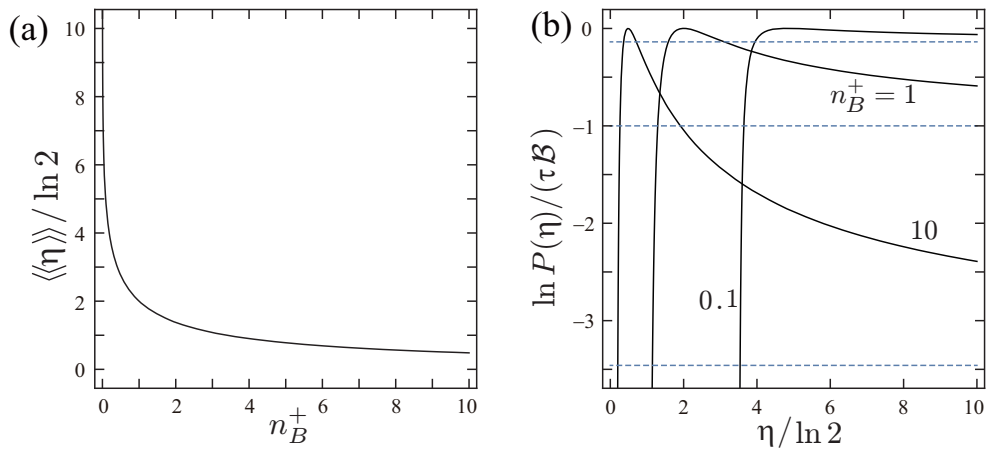} }
\end{center}
\caption{(a) The information content carried by a single boson particle. 
(b) The probability distribution of ratio $\eta =I_A/N_A$. 
Dashed lines indicate $-N \ln n_B^-$. 
}
\label{capsin}
\end{figure}

Figure \ref{capsin} (a) shows the boson occupation number $n_B^+$ dependence of the average value. 
In the limit of small boson occupation number $n_B^+ \ll 1$ it goes to infinity 
$\langle \eta \rangle \approx - \ln n_B^+$, 
as predicted in Ref.~\cite{Yamamoto1986}. 
Figure \ref{capsin} (b) shows the probability distributions for various boson occupation numbers. 
For large $\eta$, the probability decays in power law fashion 
$P_\tau (\eta) \approx (1+n_B^+)^{-N}/\eta^3$. 
This implies that, for a small boson occupation number, the information carried by a single particle fluctuates strongly. 
It can be understood by the following: 
A sequence ${\bf x}$ satisfying $N_A=\sum_{j=1}^\infty (j-1) N p_{\bf x}(a_j)$ carries the self-information $I_A=-\ln q^N({\bf x})=-N_A \ln r - N \ln (1-r)$. 
For small $n_B^+=r/(1-r) \ll 1$, the ratio  for this particular sequence is $\eta \approx -\ln n_B^+ +N n_B^+/N_A$. 
The first term is the average value and the second term is the fluctuation. 
The fluctuation is strongly enhanced for a sequence, in which the number of bosons is much smaller than the average number of signal quanta $N_A \ll N n_B^+=\tau {\mathcal P}_A/(hf)$. 

For fermions, the probability distribution of the ratio was analyzed in Ref.~\cite{YU2017}.

\section{Summary}
\label{sec:summary}

In summary, we present the fermionic and bosonic path integral for the full counting statistics and the R\'enyi entanglement entropy. 
The key relation (\ref{eqn:renyi_cgf}) holds for the quadratic action, i.e., noninteracting particles. 
We analyzed the ratio of self-information to the number of bosons transmitting through a narrow-band quantum channel. 
For the small occupation number of bosons, the average of the ratio diverges. 
At the same time, for an event in which the number of bosons is much smaller than the average number of signal quanta for a given signal power, the fluctuation of the ratio is enhanced. 

We thank Hiroaki Okada for valuable discussions. 
This work was supported by JSPS KAKENHI Grants No. 17K05575 and No. JP26220711.

\begin{appendix}

\section{Short derivation of Sanov's theorem}
\label{sec:sanov}

Let ${\mathcal X}=\{ a_1,\cdots, a_{|{\mathcal X}|} \}$ and consider the distribution $q$ on ${\mathcal X}$. 
Let $X_1,\cdots,X_N$ be a sequence of random variables drawn i.i.d. according to $q(x)$. 
The probability that the sample average of $g(X)$ is equal to $\alpha$ is, 
\begin{eqnarray}
P \left( \sum_{n=1}^N g(x_n) = N \alpha \right) = \sum_{ {\bf x} \in {\mathcal X}^N }q^N({\bf x}) \, \delta \left( \sum_{n=1}^N g(x_n)-N \alpha \right). 
\end{eqnarray}
By using the multinomial coefficient, it is written as,
\begin{eqnarray}
P
&\approx&
N^{|{\mathcal X}|}
\int d p_1 \cdots d p_{| {\mathcal X}|}
\left(
\begin{array}{ccc}
 & N & \\
N p_1 & \cdots & N p_{| {\mathcal X}|}
\end{array}
\right) q(a_1)^{N p_1} \cdots q(a_{ | {\mathcal X}| })^{N p_{ | {\mathcal X}|} }
\nonumber \\
&&\times
\delta \left( N \left( \sum_{j=1}^{|{\mathcal X}|} p_j g(a_j)-\alpha \right) \right) \delta \left( N \left( \sum_{j=1}^{|{\mathcal X}|} p_j -1 \right) \right)
\, .
\end{eqnarray}
By rewriting the delta function as, e.g., $\delta(x) =\int d \xi e^{-i \xi x}/(2 \pi)$ and by using Stirling's approximation, we obtain 
$ P \approx \exp \left(-N \min_{ \{p_j \} ,i \xi, i \lambda \in {\mathbf R}}  J(p, \xi,\lambda) \right)$ 
within the saddlepoint approximation, 
where 
$J = D(p \| q) +  i \xi \left( \sum_{j=1}^{|{\mathcal X}|} p_j g(a_j)-\alpha \right) + i \lambda  \left( \sum_{j=1}^{|{\mathcal X}|} p_j-1 \right)$. 
For the precise derivation, see Ref.~\cite{Cover2006}. 

The result is rephrased as 
$ P \approx e^{-N D(p^* \| q) }$
where $p^*$ is the closest to $q$ in relative entropy under the constraint 
$\sum_{j=1}^{|{\mathcal X}|} p_j g(a_j) =\alpha$
and
$\sum_{j=1}^{|{\mathcal X}|} p_j=1$.
For our problems, we set $a_j=j-1$ and $|{\mathcal X}|=2 (\infty)$ for fermions (bosons). 
Equation~(\ref{eqn:pdf_n}) is obtained by setting $\alpha=N_A/N$ and $g(a_j)=j-1$. 
Equation~(\ref{eqn:pdf_I}) is obtained by setting $\alpha=I_A/N$ and $g(a_j)=-\ln q(a_j)$.

\section{Path integral on multiple Keldysh contours}
\label{sec:tst}

\begin{figure}
\begin{center}
\resizebox{0.5 \columnwidth}{!}{ \includegraphics{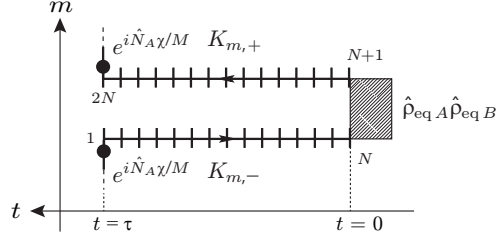} }
\end{center}
\caption{Discrete time points on the $m$th Keldysh contour. }
\label{tsmkc}
\end{figure}

We adopt the time-slicing technique~\cite{Kamenev2011,Tang2014,SeoaneSouto2015,SeoaneSouto2017,YU2002} to introduce the path-integral representation of Eq.~(\ref{eqn:mrenyi}); 
$S_M(\chi) = {\rm Tr}_A \left[ \left( e^{ i \hat{N}_A \chi/M} \hat{\rho}_A (\tau) \right)^M \right]$. 
This is the characteristic function for $M=1$, $S_1(\chi) = {\mathcal Z}_\tau(e^{i \chi})$ Eq.~(\ref{eqn:pdfnq}), and the R\'enyi entanglement entropy for $\chi=0$, $S_M=S_M(0)$ Eq.~(\ref{eqn:pdfiq}). 
Initially, when the two subsystems are decoupled, it is 
$s_M(\chi) ={\rm Tr}_A \left[ \left( e^{ i \hat{N}_A \chi/M} \hat{\rho}_{{\rm eq} A} \right)^M \right] = Z(M \beta_A, \mu_A+i \chi/(M \beta_A))/ Z(\beta_A, \mu_A)^M$. 
%
For simplicity, we consider one quantum state in each subsystem, and thus the Hamiltonians are, 
$\hat{H}_A = \epsilon_A \hat{a}^\dagger \hat{a}$, 
$\hat{H}_B = \epsilon_B \hat{b}^\dagger \hat{b}$, 
and
$\hat{V} = J ( \hat{a}^\dagger \hat{b} + \hat{b}^\dagger \hat{a} )$. 
To obtain the path-integral representation, we discretize each branch of $M$ Keldysh contours into $N-1$ time steps [Fig.~\ref{tsmkc}]. 
The step size is $d \tau=\tau/(N-1)$. A discrete time on the lower branch of the $m$th Keldysh contour is $t_{m,j}=(N-j) d \tau$ for $j=1,\cdots, N$ and that on the upper branch is $t_{m,j}=(j-N-1) d \tau$ for $j=N+1,\cdots,2N$. 
Then, at time $t_{m,j}$, we insert the closure relation, e.g.,
 
$\hat{1}=\int d a_{m,j}^* d a_{m,j} e^{-a_{m,j}^* a_{m,j}} |a_{m,j} \rangle \langle a_{m,j}|/{\mathcal N}$, 
where ${\mathcal N}=1 (2 \pi i)$ for fermions (bosons) (we follow the convention of the textbook~\cite{Negele1988}) and $\hat{a}^\dagger |a_{m,j} \rangle = a_{m,j} |a_{m,j} \rangle$ is the coherent state. 

By inserting closure relations for subsystem $A$ at $\tau_{m,+}=t_{m,2N}$ and $\tau_{m,-}=t_{m,1}$ and using the trace expression, 
${\rm Tr}_A \hat{ {\mathcal O} } = \int d a_{1,1}^* d a_{1,1} e^{-a_{1,1}^* a_{1,1}} \langle e^{-i \phi} a_{1,1}| \hat{ {\mathcal O} } |a_{1,1} \rangle / {\mathcal N}$, 
the R\'enyi entropy becomes, 
\begin{eqnarray}
S_M(\chi) 
&=& {\rm Tr}_A \left[ 
e^{ i \hat{N}_A \chi/M} \hat{\rho}_A (\tau) \cdots e^{ i \hat{N}_A \chi/M} \hat{\rho}_A (\tau) 
\right]
=
\int \frac{d a_{M,2N}^* d a_{M,2N}}{ {\mathcal N} } \int \frac{d a_{M,1}^* d a_{M,1}}{ {\mathcal N} } \nonumber \\ && \cdots  
\int \frac{d a_{1,2N}^* d a_{1,2N}}{ {\mathcal N} } \int \frac{d a_{1,1}^* d a_{1,1}}{ {\mathcal N} } e^{-\sum_{m=1}^M (a_{m,2N}^* a_{m,2N}+a_{m,1}^* a_{m,1})} 
\nonumber \\ && \times \langle e^{-i \phi} a_{1,1}| e^{i \hat{N}_A \chi/M} | a_{M,2N} \rangle \langle a_{M,2N}| \hat{\rho}_A(\tau) | a_{M,1} \rangle \langle a_{M,1}| e^{i \hat{N}_A \chi/M} | a_{M-1,2N} \rangle
\cdots \nonumber \\ && \times
\langle a_{2,2N}| \hat{\rho}_A(\tau) | a_{2,1} \rangle \langle a_{2,1}| e^{i \hat{N}_A \chi/M} | a_{1,2N} \rangle \langle a_{1,2N}| \hat{\rho}_A(\tau) | a_{1,1} \rangle
\, . 
\end{eqnarray}
Then by using the relation, 
$\langle a| e^{i \hat{N}_A \chi/M} |a^\prime \rangle = e^{a^* a^\prime e^{i \chi/M} }$, 
we obtain, 
\begin{eqnarray}
S_M(\chi) = \prod_{m=1}^M
\int \frac{ d a_{m,2N}^* d a_{m,2N} d a_{m,1}^* d a_{m,1} }{ {\mathcal N}^2 }
\rho_A(a_{m,2N},a_{m,1}; \tau)
e^{ \sum_{m=1}^M i {\mathcal S}_{{\rm b.c.} A}^m}
\, ,
\end{eqnarray}
The action, 
$i {\mathcal S}_{ {\rm b.c.} A}^m = a_{m+1,1}^* \left( a_{m,2N} e^{i \chi/M} - a_{m+1,1} \right)$
where $a_{M+1,1}=e^{-i \phi} a_{1,1}$, 
corresponds to Eq.~(\ref{eqn:bcA}) and imposes the boundary condition of the field $a$ at $\tau_{m,+}$ and $\tau_{m+1,-}$. 
$\rho_A(a_{m,2N},a_{m,1}; \tau)=\langle a_{m,2N}| \hat{\rho}_A(\tau) | a_{m,1} \rangle  e^{-a_{m,2N}^* a_{m,2N}}$ 
is the reduced density matrix expressed by the double path-integral~\cite{Schoen1990,Weiss2012,Feynman2010}. 
\begin{eqnarray}
\rho_A(a_{2N},a_{1}; \tau)
=
\prod_{j=2}^{2N-1} \int \frac{ d a_j^* d a_j }{ {\mathcal N} } \prod_{j=1}^{2N} \int \frac{ d b_j^* d b_j }{ {\mathcal N} } \frac {e^{ i {\mathcal S}_K + i {\mathcal S}_{\rm b.c.} (b_{1(2N)},b_{1(2N)}^*;\phi) } }{Z_A Z_B}
,
\end{eqnarray}
where we omitted the replica index $m$. 
The action corresponding to Eq.~(\ref{eqn:action_bulk}) is, 
\begin{eqnarray}
i {\mathcal S}_K 
& \approx &
\left( \sum_{j=N+1}^{2N-1} + \sum_{j=1}^{N-1} \right) \biggl [-a_{j+1}^* (a_{j+1} - a_j) -b_{j+1}^* (b_{j+1} - b_j)
\nonumber \\ && - i \, {\rm sgn}(j-N) \, H(a_{j+1}^*,b_{j+1}^*,a_j,b_j) \, d \tau \biggl ] \nonumber \\ && 
+ a_{N+1}^* \left( a_{N} e^{-\beta_A (\epsilon_A - \mu_A)} - a_{N+1} \right) + b_{N+1}^* \left( b_{N} e^{-\beta_B (\epsilon_B - \mu_B)} - b_{N+1} \right)
.
\end{eqnarray}
The action 
$i {\mathcal S}_{\rm b.c.}(b_{m,1(2N)},b_{m,1(2N)}^*;\phi) = b_{m,1}^* \left( b_{m,2N} e^{i \phi} - b_{m,1} \right)$ 
corresponds to Eq.~(\ref{eqn:action_boundary_condition}) and imposes the boundary condition of the field $b$ at $\tau_{m,\pm}$. 
Summarizing above, 
\begin{eqnarray}
S_M(\chi) &=& \lim_{N \to \infty} \prod_{m=1}^{M} \prod_{j=1}^{2N} \int \frac{d a_{m,j}^* d a_{m,j}}{ {\mathcal N} } \int \frac{d b_{m,j}^* d b_{m,j}}{ {\mathcal N} } e^{i {\tilde {\mathcal S}} }
\, ,
\end{eqnarray}
where the action is, 
\begin{eqnarray}
{\tilde {\mathcal S} }
= 
\sum_{m=1}^M 
{\mathcal S}_{K_m}( \{ a_{m,j} (b_{m,j}),a_{m,j}^* (b_{m,j}^*) \}) + {\mathcal S}_{\rm b.c.}( b_{m,1(2N)}, b_{m,1(2N)}^* ; \phi) + {\mathcal S}_{{\rm b.c.} A}^m 
\, .
\end{eqnarray}

In the following, we perform the functional integral. 
The matrix form of the action is, 
\begin{eqnarray}
{\tilde {\mathcal S} } = {\bf a}^\dagger {\bf g}_{A,M}^{-1} {\bf a} + {\bf b}^\dagger \left( {\bf 1}_M \otimes {\bf g}_{B,\phi}^{-1} \right) {\bf b}
-i J d \tau \left[ {\bf a}^\dagger \left( {\bf 1}_M \otimes {\bf P} \right)  {\bf b} + {\bf b}^\dagger \left( {\bf 1}_M \otimes {\bf P} \right) {\bf a} \right], 
\end{eqnarray}
where ${\bf 1}_M$ is $M \times M$ identity matrix 
and $\otimes$ stands for the Kronecker product. 
Vectors ${\bf a}$ and ${\bf b}$ consist of $2NM$ components of fields, 
e.g.,
${\bf a}^T =({\bf a}_1^T,\cdots,{\bf a}_M^T)$ 
where
${\bf a}_m =(a_{m,1},\cdots,a_{m,2N})^T$. 
The inverse Green function of subsystem $A$ is a block (skew) circulant for $\phi=0$ ($\phi=\pi$) and that of subsystem $B$ is block diagonal. 
They are, e.g., for $M=3$,
\begin{eqnarray}
i {\bf g}_{A,M}^{-1} &=&
\left[
\begin{array}{ccc}
{\bf X}_A & {\bf 0} & e^{i \chi/M+i \phi} {\bf Y}  \\
e^{i \chi/M} {\bf Y} & {\bf X}_A & {\bf 0} \\
{\bf 0} & e^{i \chi/M} {\bf Y} & {\bf X}_A 
\end{array}
\right], 
\\
{\bf 1}_M \otimes i {\bf g}_{B,\phi}^{-1} 
&=&
\left[
\begin{array}{ccc}
i {\bf g}_{B,\phi}^{-1} & {\bf 0} & {\bf 0} \\
{\bf 0} & i {\bf g}_{B,\phi}^{-1} & {\bf 0} \\
{\bf 0} & {\bf 0} & i {\bf g}_{B,\phi}^{-1}
\end{array}
\right], 
\;\;
i {\bf g}_{B,\phi}^{-1} = {\bf X}_B+e^{i \phi} {\bf Y}
\, . 
\end{eqnarray}
$2 N \times 2 N$ submatrices are, 
${\bf P}={\bf \tau}_3 \otimes {\bf p}_{N,-}$, 
${\bf Y}={\bf \tau}_+ \otimes {\bf p}_{N,+}$
and 
\begin{eqnarray}
{\bf X}_r
=
{\bf \tau}_0 \otimes (-{\bf 1}_N + {\bf p}_{N,-})
+
{\bf \tau}_+^\dagger \otimes {\bf p}_{N,+} e^{-\beta_r(\epsilon_r - \mu_r)}
+
i d \tau \epsilon_r 
{\bf \tau}_3 \otimes {\bf p}_{N,-}
\, , 
\end{eqnarray}
where ${\bf \tau}_{+} = {\bf p}_{2,+}$ is a $2 \times 2$ matrix. 
${\bf p}_{N,\pm}$ are $N \times N$ matrices, and their $(i,j)$-components are 
$[ {\bf p}_{N,+} ]_{ij}=\delta_{i,1} \delta_{j,N}$ and $[ {\bf p}_{N,-} ]_{ij}=\delta_{i,j+1}$. 
Explicit forms are, e.g., for $N=3$, 
\begin{eqnarray}
{\bf p}_{N,+}
=
\left[
\begin{array}{ccc}
0 & 0 & 1 \\
0 & 0 & 0 \\
0 & 0 & 0 
\end{array}
\right] , 
\;\;
{\bf p}_{N,-}
=
\left[
\begin{array}{ccc}
0 & 0 & 0 \\
1 & 0 & 0 \\
0 & 1 & 0 
\end{array}
\right]
\, .
\end{eqnarray}
%
The block (skew) circulant matrix ${\bf g}_{A,M}^{-1}$ can be block-diagonalized by using the discrete Fourier transform corresponding to Eq.~(\ref{eqn:ft}), 
${\bf a}_m = \sum_{\ell=0}^{M-1} {\bf a}_\ell e^{-i (2 \pi \ell+\phi)m/M}/\sqrt{M}$. 
Then, by introducing $2N \times 2N$ sub-matrix 
$i {\bf g}_{A,\chi_\ell+\chi/M+\phi}^{-1}={\bf X}_A+ e^{i (2 \pi \ell+\phi+\chi)/M} {\bf Y}$, the action becomes, 
\begin{eqnarray}
i {\tilde {\mathcal S}} = \sum_{\ell=0}^{M-1}
\left[ {\bf a}_\ell^\dagger i {\bf g}_{A,\chi_\ell+\chi/M+\phi}^{-1} {\bf a}_\ell + {\bf b}_\ell^\dagger i {\bf g}_{B,\phi}^{-1} {\bf b}_\ell -i 
J d \tau \left( {\bf a}_\ell^\dagger {\bf P} {\bf b}_\ell + {\bf b}_\ell^\dagger {\bf P} {\bf a}_\ell \right) \right]
\, . 
\end{eqnarray}
This action corresponds to Eq.~(\ref{eqn:cgf_action}) for $M=1$ and to Eq.~(\ref{eqn:diag_action}) for $\chi=0$. 
Since the Jacobian of the discrete Fourier transform is $1$, 
by exploiting the Gauss integral, 
$\ln \int \prod_{j=1}^{2N} (d a_j^* d a_j/ {\mathcal N}) e^{- {\bf a}^\dagger {\bf M} {\bf a} } = -e^{i \phi} \ln {\rm det} {\bf M}$, 
where ${\bf M}$ is a $2N \times 2N$ matrix, 
the R\'enyi entropy is calculated as, 
\begin{eqnarray}
\ln 
\frac{S_M(\chi)}{s_M(\chi)} = -e^{i \phi} \sum_{\ell=0}^{M-1} \ln \det \left[ {\bf 1}_{2N} - (J d \tau)^2 {\bf g}_{B,\phi}  {\bf P} {\bf g}_{A,\chi_\ell+\chi/M+\phi} {\bf P} \right] \, . 
\label{eqn:renyi_mtc}
\end{eqnarray}
Here the Green function is, 
\begin{eqnarray}
\left[ {\bf g}_{r,\lambda} \right]_{ij} = -i 
\left \{
\begin{array}{cc}
n_{r,\lambda}^-
e^{-i \epsilon_r (j-i) d \tau} & (1 < j < i < N ) \\
n_{r,\lambda}^+ 
e^{-i \lambda + i \epsilon_r (2N-i-j+1) d \tau}  & (1 < j < N, N+1 \leq i \leq 2N ) \\
n_{r,\lambda}^-
e^{- i \epsilon_r (i-j) d \tau} & (N+1 < j < i < 2 N ) \\
n_{r,\lambda}^+ e^{-i \epsilon_r (j-i) d \tau} & (1 < i < j < N ) \\
n_{r,\lambda}^- e^{i \lambda -i \epsilon_r (2N -i-j+1) d \tau}  & (1 < i < N, N+1 < j < 2N ) \\
n_{r,\lambda}^+ e^{-i \epsilon_r (i-j) d \tau} & (N+1 < i < j < 2 N )
\end{array}
\right. 
\, . 
\end{eqnarray}
In the time-continuous limit $d \tau \to 0$, we set $t=t_i$ and $t'=t_j$. 
It is $t_i = (N-i) d \tau \in K_-$ for $1 \leq i \leq N$ and $t_i = (i-N-1) d \tau \in K_+$ for $N+1 \leq i \leq 2N$. 
Then, in the $2 \times 2$ Keldysh Green function matrix form, 
\begin{eqnarray}
{\bf g}_{r,\lambda}(t,t')
&=&
\left [
\begin{array}{cc}
g_{r,\lambda}(t \in K_+,t' \in K_+) &  g_{r,\lambda}(t \in K_+,t' \in K_-) \\
g_{r,\lambda}(t \in K_-,t' \in K_+) &  g_{r,\lambda}(t \in K_-,t' \in K_-)
\end{array}
\right]
=
-i e^{-i \epsilon_r (t-t')}
\nonumber \\
&& \times
\left [
\begin{array}{cc}
n_{r,\lambda}^- \theta(t-t') + n_{r,\lambda}^+ \theta(t'-t)
& 
n_{r,\lambda}^+ e^{-i \lambda} \\
n_{r,\lambda}^- e^{ i \lambda} 
&
n_{r,\lambda}^- \theta(t'-t) + n_{r,\lambda}^+ \theta(t-t')
\end{array}
\right]
.
\label{app:mkgf}
\end{eqnarray}
In the limit of $\tau \to \infty$, the logarithm in Eq.~(\ref{eqn:renyi_mtc}) is expanded as,  
\begin{eqnarray}
\ln \det[\cdots] &=& - J^2 \int_0^\tau dt_2 dt_1
{\rm Tr} \left[ {\bf \tau}_{3} {\bf g}_{B,\phi}(t_1,t_2) {\bf \tau}_3 {\bf g}_{A,\chi_\ell+\chi/M+\phi}(t_2,t_1) \right] + \cdots  
\nonumber \\
& \approx &
- \frac{\tau}{2 \pi} J^2 \int d \omega {\rm Tr} \left[ {\bf \tau}_{3} {\bf g}_{B,\phi}(\omega) {\bf \tau}_3 {\bf g}_{A,\chi_\ell+\chi/M+\phi}(\omega) \right] + \cdots
\nonumber \\
&=&
\frac{\tau}{2 \pi} \int d \omega {\rm Tr} \ln 
\left[ {\bf \tau}_{0} - J^2 
{\bf \tau}_3 {\bf g}_{A,\chi_\ell+\chi/M+\phi}(\omega) 
{\bf \tau}_{3} {\bf g}_{B,\phi}(\omega) 
\right]
\, ,
\label{eqn:f}
\end{eqnarray}
where the Fourier transform of Eq.~(\ref{app:mkgf}) is, 
\begin{eqnarray}
{\bf g}_{r,\lambda}(\omega) = {\rm P}. \frac{1}{\omega - \epsilon_r} \, {\mathbf \tau}_3 - 2 \pi i \, \delta (\omega - \epsilon_r)
\left [
\begin{array}{cc}
1/2+n_{r,\lambda}^+(\omega) & n_{r,\lambda}^+(\omega) e^{-i \lambda} \\
n_{r,\lambda}^-(\omega) e^{ i \lambda} & 1/2+n_{r,\lambda}^+(\omega)
\end{array}
\right]
\, . 
\end{eqnarray}
The above calculations can be readily extended to many states $\epsilon_r \to \epsilon_{rk}$, and then Eq.~(\ref{eqn:kgm}) is 
${\bf g}_{r,\lambda}(\omega) = \sum_k {\bf g}_{r k, \lambda}(\omega)$. 
By combining Eqs. (\ref{eqn:renyi_mtc}) and (\ref{eqn:f}), we obtain, 
\begin{eqnarray}
\ln \frac{S_M(\chi)}{s_M(\chi)} \approx -e^{i \phi} \sum_{\ell=0}^{M-1} \frac{\tau}{2 \pi} \int d \omega {\rm Tr} \ln 
\left[ {\bf \tau}_{0} - J^2 {\bf \tau}_3 {\bf g}_{A,\chi_\ell+\chi/M+\phi}(\omega) {\bf \tau}_{3} {\bf g}_{B,\phi}(\omega) 
\right]
\, , 
\end{eqnarray}
which corresponds to Eq.~(\ref{eqn:cgf_omega}) for $M=1$ and to Eq.~(\ref{eqn:renyi_cgf}) for $\chi=0$.

\end{appendix}


\begin{thebibliography}{99}


\bibitem{Blanter2000}
Ya.~M.~Blanter, and M.~B\"uttiker, 
Phys. Rep. {\bf 336} (2000) 1.

\bibitem{Levitov1996}
L. S. Levitov, H. W. Lee, and G. B. Lesovik, 
J. Math. Phys. {\bf 37} (1996) 4845. 

\bibitem{Nazarov2003}
{\it Quantum Noise in Mesoscopic Physics, 
Vol. 97 of NATO Science Series II: Mathematics, Physics and Chemistry} 
edited by Yu. V. Nazarov (Kluwer Academic Publishers, Dordrecht/Boston/London, 2003).

\bibitem{BUGS2006}
D.~A.~Bagrets, Y.~Utsumi, D.~S.~Golubev, and Gerd~Sch\"on, 
Fortschritte der Physik {\bf 54} (2006), 917-938. 

\bibitem{Hogg2005}
See, e.g.,
R. V. Hogg, J. W. McKean, and A. T. Craig, 
{\it Introduction to Mathematical Statistics}, 6th ed. 
(Pearson Education, Upper Saddle River, New Jersey, 2005). 

\bibitem{Bennakker2006}
C. W. J. Beenakker, 
{\it Proceedings of the International School of Physics Enrico Fermi}, 
Vol. 162 (IOS Press, Amsterdam, 2006). 


\bibitem{Shannon1948}
C.~E.~Shannon, 
Bell System Techn. J. {\bf 27} (1948) 379-423, 623-656.

\bibitem{Cover2006}
T. M. Cover and J. A. Thomas, {\it Elements of Information Theory}, 2nd ed. (Wiley-Interscience, New York, 2006). 

\bibitem{NC2000}
M. A. Nielsen and I. L. Chuang, {\it Quantum Computation and Quantum Information}, (Cambridge University Press, New York, 2000).

\bibitem{Klich2009}
I. Klich and L. S. Levitov, Phys. Rev. Lett. {\bf 102} (2009) 100502. 

\bibitem{FrancisSong2012}
H. Francis Song, S. Rachel, C. Flindt, I. Klich, N. Laflorencie, and K. Le Hur, Phys. Rev. B {\bf 85} (2012) 035409. 

\bibitem{Suesstrunk2012}
R.~S\"usstrunk and D.~A.~Ivanov, EPL {\bf 100} 60009 (2012).

\bibitem{Calabrese2012}
P.~Calabrese, M.~Mintchev, and E.~Vicari, EPL {\bf 98} 20003 (2012).


\bibitem{Hsu2009}
B.~Hsu, E.~Grosfeld, and E.~Fradkin, 
Phys. Rev. B {\bf 80}, 235412 (2009). 

\bibitem{Renyi1960} 
A. R\'enyi, in {\it Proceedings of the Fourth Berkeley Symposium on Mathematics, Statistics, and Probability} (University of California Press, Berkeley, CA, 1960), p. 547.

\bibitem{Nazarov2011}
Yu. V. Nazarov, Phys. Rev. B {\bf 84} (2011) 205437. 

\bibitem{Ansari2015}
M. H. Ansari and Yu. V. Nazarov, 
Phys. Rev. B {\bf 91} (2015) 174307; 
ZhETF, 2016, {\bf 149} (2016) 453. 

\bibitem{Ansari2017}
M. H. Ansari, arXiv:1605.04620; Phys. Rev. B {\bf 95} (2017) 174302. 

\bibitem{YU2015}
Y. Utsumi, Phys. Rev. B {\bf 92} (2015) 165312. 

\bibitem{YU2017}
Y. Utsumi, Phys. Rev. B {\bf 96} (2017) 085304. 

\bibitem{Casini2009} 
H.~Casini, and M.~Huerta, 
J.~Phys.~A, {\bf42} (2009) 504007; arXiv:0905.2562v3. 

\bibitem{Yamamoto1986}
Y.~Yamomoto and H.~A.~Haus, Rev. Mod. Phys. {\bf 58} (1986) 1001. 

\bibitem{Touchette2009}
H. Touchette, Phys. Rep. {\bf 478} (2009) 1.

\bibitem{Fano1961}
R.~M.~Fano, {\it Transmission of Information: A Statistical Theory of Communications}, (The M.I.T. Press, Cambridge, 1961).  

\bibitem{Golomb1966}
S. W. Golomb, IEEE Trans. Inform. Theory {\bf IT-12} (1966) 75. 

\bibitem{Guiasu1985}
S. Guiasu and C. Reischer, Information Sciences, {\bf 35} (1985) 235. 

\bibitem{Schoen1990}
Gerd~Sch\"on and A.~D.~Zaikin, Phys. Rep. {\bf 198} (1990) 237-412. 

\bibitem{Weiss2012}
U. Weiss, {\it Quantum Dissipative Systems}, 4th ed. (World Scientific, Singapore, 2012). 

\bibitem{Kamenev2011}
A. Kamenev, {\it Field Theory of Nonequilibrium Systems}, (Cambridge University Press, Cambridge, 2011).

\bibitem{Negele1988}
J.~W.~Negele, and H.~Orland, {\it Quantum Many-Particle Systems}, 
(Addison Wesley, Redwood City CA, 1988). 

\bibitem{Esposito2009} 
M. Esposito, U. Harbola, and S. Mukamel, Rev. Mod. Phys. {\bf 81} (2009) 1665.

\bibitem{UGS2006}
Y. Utsumi, D. S. Golubev, Gerd Sch\"on, Phys. Rev. Lett. {\bf 96} (2006) 086803. 

\bibitem{SU2008}
K. Saito and Y. Utsumi, Phys. Rev. B {\bf 78} (2008) 115429. 

\bibitem{US2009}
Y. Utsumi and K. Saito, Phys. Rev. B {\bf 79} (2009) 235311. 

\bibitem{Urban2010}
D. F. Urban, R. Avriller, and A. Levy Yeyati, Phys. Rev. B {\bf 82} (2010) 121414(R). 

\bibitem{Novotny2011}
T. Novotn\'y, F. Haupt, and W. Belzig, Phys. Rev. B {\bf 84} (2011) 113107. 

\bibitem{UEUA2013}
Y. Utsumi, O. Entin-Wohlman, A. Ueda, A. Aharony, Phys. Rev. B {\bf 87} (2013) 115407. 

\bibitem{Petrescu2014}
A. Petrescu, H. F. Song, S. Rachel, Z. Ristivojevic, C. Flindt, N. Laflorencie, I. Klich, N. Regnault, and K. Le Hur, J. Stat. Mech. (2014) P10005.

\bibitem{Jarzynski1997}
C. Jarzynski, Phys. Rev. Lett. {\bf 78} (1997) 2690. 

\bibitem{Campisi2011} 
M. Campisi, P. H\"{a}nggi, and M. Talkner, Rev. Mod. Phys. {\bf 83} (2011) 771. 

\bibitem{Abanov2009}
A. G. Abanov and D. A. Ivanov, Phys. Rev. B {\bf 79} (2009) 205315. 

\bibitem{Abanov2008}
A. G. Abanov and D. A. Ivanov, Phys. Rev. Lett. {\bf 100} (2008) 086602. 

\bibitem{Xiang-bin2002}
W.~Xiang-bin, Phys. Rev. A {\bf 66} (2002) 024303. 

\bibitem{Verley2014}
G. Verley, M. Esposito, T. Willaert, and C. Van den Broeck, Nat. Commun. {\bf 5}, 4721 (2014); G. Verley, T. Willaert, C. Van den Broeck, and M. Esposito, Phys. Rev. E {\bf 90} (2014) 052145.

\bibitem{Polettini2015}
M.~Polettini, G.~Verley, and M.~Esposito, Phys. Rev. Lett. {\bf 114} (2015) 050601.

\bibitem{Proesmans2016} K.~Proesmans, Y.~Dreher, M.~Gavrilov, J.~Bechhoefer, and C.~Van den Broeck, Phys. Rev. X {\bf 6}, 041010 (2016). 

\bibitem{Okada2017}
H.~Okada and Y.~Utsumi, J. Phys. Soc. Jpn. {\bf 86} (2017) 024007. 

\bibitem{Wiseman2003}
H. M. Wiseman and J. A. Vaccaro, Phys. Rev. Lett. {\bf 91} (2003) 097902. 

\bibitem{Klich2009_2}
I. Klich, and L. S. Levitov, arXiv:0812.0006. 

\bibitem{Tang2014}
G-M. Tang and J.~Wang, Phys. Rev. B {\bf 90} (2014) 195422. 

\bibitem{SeoaneSouto2015}
R. Seoane Souto, R. Avriller, R. C. Monreal, A. Mart\'in-Rodero, and A. Levy Yeyati, Phys. Rev. B {\bf 92} (2015) 125435. 

\bibitem{SeoaneSouto2017}
R. Seoane Souto, A. Mart\'in-Rodero, and A. Levy Yeyati, Phys. Rev. Lett. {\bf 117}, 267701 (2016); Phys. Rev. B {\bf 96} (2017) 165444. 

\bibitem{YU2002}
Y.~Utsumi, H.~Imamura, M.~Hayashi, and H.~Ebisawa, 
Phys. Rev. B {\bf 66} (2002) 024513. 

\bibitem{Feynman2010}
R.~P.~Feynman, A.~R.~Hibbs, and D.~F.~Styer, {\it Quantum Mechanics and Path Integrals: Emended Edition} (Dover, Mineola, New York, 2010). 

\end{thebibliography}
\end{document}